\documentstyle[12pt]{article}
\newcommand{\C}{{\bf C}}

\newcommand{\N}{{\bf N}}
\newcommand{\R}{{\bf R}}

\newcommand{\be}[1]{\begin{equation}\label{#1}}
\newcommand{\ee}{\end{equation}}
\newcommand{\rf}[1]{(\ref{#1})}

\newcommand{\proof}{\vspace*{2mm}\noindent{\bf Proof. }\ }
\newcommand{\eproof}{\rule{0.5em}{0.5em}}

\title{On spectra of quadratic operator pencils with rank one  gyroscopic linear part}
\author{Olga Boyko, Olga Martynyuk, Vyacheslav Pivovarchik  \ \ }

\begin{document}
\maketitle

Abstract. The spectrum of  a selfadjoint quadratic operator pencil of the form \\
$\lambda^2M-\lambda G-A$ is investigated where $M\geq 0$,  $G\geq 0$ are bounded operators and $A$   is selfadjoint bounded below
is investigated. It is shown that  in the case of rank one operator $G$ the eigenvalues of such a pencil are of two types.  The eigenvalues of one of these types are independent of the operator 
$G$.  Location of the eigenvalues of both types is described. Examples for the case of the Sturm-Liouville operators $A$ are given. 

\vspace{5mm}
keywords: quadratic operator pencil, gyroscopic force, eigenvalues, algebraic multiplicity

\vspace{5mm} \

\newpage

\section{Introduction.}
\setcounter{equation}{0}

Quadratic operator pencils of the form $L(\lambda)=\lambda^2M-\lambda G-A$ with a selfadjoint operator $A$ bounded below describing potential energy, a
bounded symmetric operator $M\geq 0$ describing inertia of the system and an operator $G$  bounded or subordinate to  $A$ occur in different physical problems where,
 in most of cases, they have spectra consisting of normal eigenvalues (see below Definition 2.2). 
Usually the operator $G$ is  symmetric (see, e.g. \cite{KO}, \cite{Sh} and Chapter 4 in \cite{MP}) or antisymmetric
 (see \cite{Pumz} and Chapter 2 in \cite{MP}). In the first case $G$ describes gyroscopic effect while in the latter case damping forces. 

The problems in which gyroscopic forces occur can be found in  \cite{Fe}, \cite{DD}, \cite{Th}, \cite{Za}, \cite{Mo},  \cite{Mi1}, \cite{Mi2}, \cite{PI}. 
The spectra of problems described in these papers  may contain complex eigenvalues what leads to instability of the corresponding dynamical system. The gyroscopic stabilization 
in terms of spectral theory means, roughly speaking, that the spectrum of an operator pencil $L(\lambda)=\lambda^2M-\lambda G-A$ with selfadjoint $A$ and symmetric $M$ and $G$ can be real while 
the spectrum of $L_0(\lambda)=\lambda^2M-A$ contains pure imaginary eigenvalues.
A sufficient condition for the pencil $L$ to have only real eigenvalues is the overdamping condition:
\begin{equation}
\label{1.1}
(Gy,y)^2+4(My,y)(Ay,y)\geq 0
\end{equation}
here and in what follows $(\cdot,\cdot)$ stands for the inner product in the corresponding Hilbert space.
 In \cite{M} pencils which satisfy (\ref{1.1}) are called hyperbolic.

In the present paper we consider the case of symmetric operator $G\geq 0$, i.e. $(Gy,y)\geq 0$ for all $y\in D(G)$. As far as we know  location of  complex eigenvalues of the corresponding operator pancil was not considered before. Up to our knowledge, also the case of rank one operator $G$ in the pencil $L(\lambda)$ was not considered. 

In Section 2 we obtain some general results on location of eigenvalues of the pencil $L$.

In  Section 3 we consider
the case of  rank one operator $G$ (problems with such operator $G$ can be found e.g. in \cite{IN}). We show that the real eigenvalues are in certain sense interlaced. In case of $A>>0$ (i.e. $A\geq \epsilon I$ for some $\epsilon>0$) this 
interlacing  is equivalent to the 'self-interlacing' described in \cite{Tya} for finite dimensional (matrix) problems.

It should be mentioned   that the spectra of  problems described by the pencil $\lambda^2M-i\lambda K-A$
 with selfadjoint $K\geq 0$, $M\geq 0$  (i.e. $(Ky,y)\geq 0$, $(My,y)\geq 0$ for all $y\in D(K)$ and $y\in D(M)$, respectively)  and selfadjoint $A$ bounded below can be described also as sets of zeros of the shifted generalized Hermite-Biehler functions of the form 
$\omega(\lambda)=P(\lambda^2)+i\lambda Q(\lambda^2)$ where $P(\lambda^2)$, $\lambda Q(\lambda^2)$ (or $\lambda P(\lambda^2)$, $\lambda^2 Q(\lambda^2)$)  are sine-type functions and   $\frac{Q(z)}{P(z)}$ is essentially positive Nevanlinna function (see \cite{PW} or \cite{MP}). 

Similarly, the spectra of problems described by the pencil  $\lambda^2M-\lambda G-A$
 with selfadjoint $G\geq 0$, $M\geq 0$ and selfadjoint $A$ bounded below can be described also as sets of zeros of the  functions of the form $\omega(\lambda)=P(\lambda^2)+\lambda Q(\lambda^2)$ where $\frac{Q(z)}{P(z)}$
 is again essentially positive Nevanlinna function (see the function $\omega$ in (\ref{yyy}) in Section 4 as an example).

 In Section 4 the results obtained in Section 3 are applied to a Sturm-Liouville boundary value problems with a boundary condition linear in the spectral parameter. Our results give 
necessary   conditions (Theorem 4.1) for a sequence of numbers to be the spectrum of such a problem (problem (\ref{4.1})--(\ref{4.3}). In our future publication we will show that these conditions are also sufficient and give a method
 of recovering the potential of the Sturm-Liouville equation. We will also show that it is enough to know one spectrum of such  a boundary value problem (problem (\ref{4.1})--(\ref{4.3}) to recover the potential of the Sturm-Liouville equation.

We give also in Section 4 an example of a boundary value problem the spectrum of which consists of two subsequences one of which  is symmetric with respect to the real and to the imaginary axes and  independent of $G$.

\section{Abstract Results}
\setcounter{equation}{0}

Let us denote by $B(H)$ the set of  bounded closed operators acting on a
separable Hilbert space $H$. We deal  with the following
quadratic operator pencil
$$
L(\lambda)=\lambda^2M-\lambda G-A,
$$
where $M\in B(H)$, $G\in B(H)$ and $A$ is a selfadjoint operator on $H$.
Since $D(M)=D(G)=H$, the domain of the pencil is
chosen as usually: $D(L(\lambda))=D(M)\cap D(G)\cap D(A)=D(A)$.
Thus, it is independent of $\lambda$.

\vspace*{1mm} In what follows we assume the following condition
to be satisfied:

\vspace*{3mm} {\bf Conditions I}: {\it $M\in B(H)$, $M\geq 0$, $G\in B(H)$ and $G\geq 0$, $A=A^{*}\geq -\beta I$ ($\beta$ is a positive
number); for some $\beta_1>\beta$ there exists $(A+\beta_1I)^
{-1}\in S_{\infty}$, where by $S_{\infty}$ we denote the set of
compact operators on $H$; $ker A\cap ker G\cap ker M=\{0\}$.}

\vspace*{3mm} {\bf Definition 2.1.} {\it The set of values
$\lambda\in\C$ such that $L^{-1}(\lambda):=L(\lambda)^{-1}$ exists in $B(H)$ is
said to be the resolvent set $\rho(L)$ of $L$.
The spectrum of the pencil $L$ is denoted by
$\sigma(L)$, i.e.
$\sigma(L)=\C\backslash\rho(L)$}.

\vspace*{3mm} {\bf Definition 2.2.} (see, e.g. \cite{MP}, Definition 1.1.3)  {\it 1. A number $\lambda_0\in\C$
is said to be an eigenvalue of the pencil $L$ if there
exists a vector $y_0\in D(A)$ (called an eigenvector of
$L$, corresponding to $\lambda_0$ ) such that $y_0\not=0$ and $L(\lambda_0)y_0=0$.
Vectors $y_1, ...,y_{m-1}$ are called associated to $y_0$ if }
\begin{equation}
\label{@}
\mathop{\sum}\limits_{s=0}^{k}\frac{1}{s!}\left.\frac{d^sL(\lambda)}{d\lambda^s}
\right|_{\lambda=\lambda_0}y_{k-s}=0, \ \ (k=1, 2, ..., m-1).
\end{equation}
{\it The number $m$ is said to be the length of the chain composed
of the eigen- and associated vectors $y_0, y_1, ..., y_{m-1}$. \\
2. The geometric multiplicity
of an eigenvalue is defined to be the number of the corresponding
linearly independent eigenvectors. 
The algebraic multiplicity of
an eigenvalue is defined to be the greatest value of the sum of
the lengths of chains corresponding to linearly independent
eigenvectors. An eigenvalue is called  semisimple if its algebraic multiplicity equals its geometric multiplicity. An eigenvalue is called simple if its algebraic multiplicity is 1. \\
3. An eigenvalue $\lambda_0$ is said to  be isolated if it has some
deleted neighborhood contained in the resolvent set $\rho(L)$. An isolated
eigenvalue $\lambda_0$ of finite algebraic multiplicity is said to
be normal. }

In the case of linear monic operator pencils $\lambda I-A$ with
bounded operator $A$ this definition of a normal eigenvalue
coincides with the corresponding definition in \cite{GK1} (Chapter
I, paragraph 2) for operators.

\vspace*{3mm} {\bf Lemma 2.3} {\it 1. The operator  pencil $L$ is Fredholm valued with index 0.\\
2. $L(\lambda)^*=L(\overline{\lambda})$ for all $\lambda\in C$. In particular, the spectrum of $L$ is symmetric with respect to the real axis.}

The proof of this lemma is the same as the  proof of   Lemma 1.1.11      in \cite{MP}.

\vspace*{3mm} {\bf Lemma 2.4} {\it 
1. The spectrum of the pencil $L$ consists of normal eigenvalues. \\
2. All (if any) the nonreal eigenvalues of
$L$ are located in the closed right
half-plane.\\
3. If $G>0$, i.e. $(Gy,y)>0$ for all $y\in D(G)$, $y\not=0$,  then all (if any) nonreal eigenvalues of $L$ are located
in the open right half-plane.}\\
 {\it 4. If $A\geq 0$, then the
spectrum $\sigma(L)$ of $L$ (if not empty)
is located on the real axis. \\
5. If $A>>0$, i.e. $A\geq\epsilon I$, $\epsilon>0$, then
 $\sigma (L)\subset \R\backslash \{0\}$. }

\proof Since the pencil $L$ is an analytic Fredholm operator valued function, its spectrum consists of eigenvalues of finite algebraic multiplicity and either 
$\sigma(L)=\C$ or all the eigenvalues are normal, see e.g., \cite{GGK}, Chapter XI, Corollary 8.4.

Let $y_0$ be an eigenvector corresponding to the eigenvalue
$\lambda_0$. Then
$$
(L(\lambda_0)y_0,y_0)=0, \ \ \ (y_0\not=0),
$$
where $(\cdot,\cdot)$ is the inner product in $H$ and consequently,
\begin{equation}
\label{2.1}
((\mbox{Re}\lambda_0)^2-(\mbox{Im}\lambda_0)^2)(My_0,y_0)-\mbox{Re}\lambda_0
(Gy_0,y_0)-(Ay_0,y_0)=0
\end{equation}
and
\begin{equation} \label{2.2}
\mbox{Im}\lambda_0(2\mbox{Re}\lambda_0(My_0,y_0)-(Gy_0,y_0))=0.
\end{equation}
If $\mbox{Im}\lambda_0\not=0$, then in view of Condition I the inequality
$\mbox{Re}\lambda_0\geq 0$ follows from (\ref{2.2}).
Hence $\sigma(L)=\C$ is impossible and assertions 1 and 2 are proved.

  If $G>0$  and 
$\mbox{Im}\lambda_0\not=0$, then (\ref{2.2}) implies $\mbox{Re}\lambda_0\not= 0$ and therefore
$\mbox{Re}\lambda_0> 0$. Assertion 3 is proved.

Now let $A\geq 0$. If $\mbox{Im}\lambda_0\not=0$ then multiplying (\ref{2.2}) by $\frac{\mbox{Re}\lambda_0}{\mbox{Im}\lambda_0}$ we subtract the resulting equation from (\ref{2.1}) and obtain
$$
- ((\mbox{Re}\lambda_0)^2+(\mbox{Im}\lambda_0)^2)(My_0,y_0)-(Ay_0,y_0)=0.
$$
This implies $(My_0,y_0)=(Ay_0,y_0)=0$.
Since the operators $M$ and $A$ are selfadjoint we obtain $My_0=Ay_0=0$, and consequently $Gy_0=0$. This contradicts Condition I. Thus, statement 4 is proved.

If $A>>0$ then in addition to the previous arguments we have $0\notin \sigma(L)$. That means statement 5 is valid.   
\eproof

\vspace*{3mm} {\bf Lemma 2.5} {\it 1. All the eigenvalues of
$L$ located  on $(-\infty,0)$ are semisimple, i.e. they do not possess associated vectors.\\
2. If $G>0$ on $ker A$, then all the eigenvalues of $L$ located on
$(-\infty,0]$ are semisimple.}

\proof 1.  Let $\lambda_0$ be an eigenvalue of $L$ located on
$(-\infty,0)$. Let us denote by $y_0$ (one of) the corresponding
eigen- and by $y_1$ the first associated vector of the chain. By Definition 2.2
\begin{equation}
\label{2.3}
\lambda_0^2My_1- \lambda_0Gy_1-Ay_1+2\lambda_0My_0-Gy_0=0.
\end{equation}
Multiplying (\ref{2.3}) by $y_0$ we obtain
\begin{equation}
\label{2.4}
((\lambda_0^2- \lambda_0G-A)y_1,y_0)+((2\lambda_0M- G)y_0,y_0)=0.
\end{equation}
Taking into account that $\lambda_0$ is real we obtain
from (\ref{2.4}):
$$
(y_1,(\lambda_0^2M- \lambda_0G-A)y_0)+((2\lambda_0M- G)y_0,y_0)=0,
$$
what means
\begin{equation}
\label{2.5}
((2\lambda_0M-G)y_0,y_0)=0.
 \end{equation}
Equation (\ref{2.5}) is possible for $\lambda_0<0$ only
if $(My_0,y_0)=(Gy_0,y_0)=0$. Since $M$ and $G$ are selfadjoint we arrive at $My_0=Gy_0=0$ and, consequently, $L(\lambda_0)y_0=-Ay_0=0$. Thus, 
$y_0\in\mbox{ker}M\cap \mbox{ker}G\cap\mbox{ker}A$. Then due to
Conditions I we have $y_0=0$, a contradiction.

2. Let $G>0$ on $ker A$ and let $y_0\not=0$ be an eigenvector corresponding to the eigenvalue $\lambda_0=0$ and therefore $y_0\in ker A$. Then (\ref{2.5}) implies $(Gy_0,y_0)=0$, a contradiction.
\eproof

{\bf Lemma 2.6} {\it
If  $M>>0$ (i.e. $(My,y)\geq m ||y||^2$, $m>0$) then all (if any) the
algebraically nonsimple real  eigenvalues  lie on the interval $[0,\frac {1}{2m}||G^{\frac{1}{2}}||^2]$.}

\proof
This result follows immediately from (\ref{2.5})
\eproof


\vspace*{3mm} {\bf Lemma 2.7} {\it If $M+G\geq\epsilon I$
($\epsilon>0$), $\mbox{dim}\mbox{ker}A>0$ and
$\mbox{dim}(\mbox{ker}A\cap\mbox{ker}G)=p\geq 0$, then the
algebraic multiplicity of $\lambda=0$ as an eigenvalue of
$L$ is equal to $p+ \mbox{dim ker}A$}.

\proof Let $0\not=y_0\in\mbox{ker}A$ and let $y_1$ be the first
associated  vector of the chain. Then
\begin{equation}
\label{2.8}
\left.\frac{dL(\lambda)}{d\lambda}\right|_{\lambda=0}y_0+L(0)y_1=
-Gy_0-Ay_1=0. \end{equation}
 If $y_0\in\mbox{ker}G$ then $y_1$
can be chosen equal to 0. If $y_0\not\in\mbox{ker}G$, then
(\ref{2.8}) implies
$$
-(Gy_0,y_0)-(Ay_1,y_0)=-(Gy_0,y_0)-(y_1,Ay_0)=-(Gy_0,y_0)=0.
$$
Combining the last equation with the condition $G\geq 0$ we obtain
$Gy_0=0$, a contradiction. It remains to prove that in case of $y_0\in ker G$, $y_1=0$ the third
vector of the chain (see Definition 2.2) does not exist. Suppose it does exist and
denote it by $y_2$. Then, see (\ref{@}),
$$
-Ay_2- Gy_1+My_0= - Ay_2+My_0=0.
$$
Consequently,
$$
0=-(Ay_2,y_0)+(My_0,y_0)=(My_0,y_0).
$$
Then  $My_0=0$ and $My_0=Gy_0=Ay_0=0$ what contradicts Condition I.
 \eproof

Let us introduce the following parameter-dependent operator
pencil: 
$$
L(\lambda,\eta)=\lambda^{2}M-\lambda\eta G-A.
$$
It is clear that $L(\lambda,1)=L(\lambda)$ and  $L(\lambda,0)=\lambda^2M-A$.

\vspace*{3mm} {\bf Theorem 2.8} (see \cite{MP}, Theorem 1.2.7). {\it Let $\eta_0\in  \C$ and let the domain $\Omega\subset \C$
be a domain which contains exactly one eigenvalue $\lambda_0$ of the pencil
$L(\lambda,\eta_0)$. Denote by $m$ the algebraic multiplicity of
$\lambda_0$. Then there exist numbers $\epsilon>0$ and $m_1\in\N$
$m_1\leq m$, such that the following assertions are true in the deleted
neighborhood
$0<|\eta-\eta_0|<\epsilon$: \\
1. The pencil $L(\lambda ,\eta)$ possesses exactly $m_1$ distinct
eigenvalues inside the domain $\Omega$. Those eigenvalues can be
arranged in groups $\lambda_{sj} (\eta)$, ($s=\overline{1,l}$;
$j=\overline{1,p_s}$; $\mathop{\sum}\limits _{s=1}^{l}p_s=m_1$) in
such a way that the functions of the group, i.e. $\lambda_{s1},
\lambda_{s2},...,\lambda_{sp_s}$ compose a complete set of a
$p_s$-valued function. In this case those eigenvalues can be
presented in the form of the following Puiseux series expansion
$$
\lambda_{sj}(\eta)=\lambda_0+\mathop{\sum}\limits_{k=1}^{\infty}a_{sk}
(((\eta-\eta_0)^{\frac{1}{p_s}})_j)^k, \ \ (j=1,2, ..., p_s),
$$
holds,  where, for $j=1, 2, ..., p_s$
$$
((\eta-\eta_0)^{\frac{1}{p_s}})_j=|\eta-\eta_0|^{\frac{1}{p_s}}exp\left(\frac{2\pi(j-1)+i\arg(\eta-\eta_0)}{p_s}\right).
$$ 

2. A basis of the eigen-space corresponding to
$\lambda_{sj}(\eta)$ can be presented in the following form
$$
y_{sj}^{(q)}(\eta)=y_{s0}^{(q)}+\mathop{\sum}\limits_{k=1}^{\infty}y_{sk}^{(q)}
(((\eta-\eta_0)^{\frac{1}{p_s}})_j)^k, \ \ (j=1,2, ...,p_s), \ \
(q=1,2, ..., \alpha_s),
$$
 where $\alpha_s$ is
the geometric multiplicity of $\lambda_{sj}(\eta)$, $y_{s0}^{(q)}$
belong to the eigen-subspace of $L(\lambda,\eta_0)$ corresponding
to the eigenvalue $\lambda_0$.}

It should be mentioned that Theorem 2.8 is a generalization of
the Weierstrass theorem on function analytic in two variables
(see, e.g. \cite{Mh} p. 476). It was proved in \cite{GS} for bounded operators. 

{\bf Remark 2.9} If $\lambda_0\not=0$ is a real or pure imaginary eigenvalue of $L(\cdot, \eta_0)$ for some real $\eta_0$ and if eigenvalues 
of $L(\cdot,\eta)$ near $\lambda_0$ for real $\eta$ near $\eta_0$ are also real or pure imaginary, then $p_s=1$ for all $s$ in Theorem 2.8.

{\bf Remark 2.10}  If both $\eta$ and $\eta_0$ are real so that $\arg(\eta-\eta_0)$ is an integer multiple of $\pi$, then we can put 
\begin{equation}
\label{2.12}
((\eta-\eta_0)^{\frac{1}{p_s}})_j=|\eta-\eta_0|^{\frac{1}{p_s}}exp\left(\frac{2\pi(j-1)+i\delta_i\arg(\eta-\eta_0)}{p_s}\right)
\end{equation}
where $\delta_i$ is an odd integer. This would change the numeration of the roots for $\eta-\eta_0<0$ if $\frac {\delta_i-1}{p_s}$ is not 
even integer but allows us when taking $\delta_s=p_s$ if $p_k$ is odd to have $((\eta-\eta_0)^{\frac{1}{p_s}})_1$ real for all $\eta$ in a real neighborhood of $\eta_0$.

\vspace*{3mm} 
If we additionally assume that $G$ is bounded and boundedly invertible then we can write, assuming $\lambda\not=0$,
$$
L(\lambda,\eta)=\lambda G^{\frac{1}{2}}(-\eta I+\lambda G^{-\frac{1}{2}}MG^{-\frac{1}{2}}-\lambda^{-1}G^{-\frac{1}{2}}AG^{-\frac{1}{2}})G^{\frac{1}{2}}.
$$ 
Hence
$$
L(\lambda,\eta)=\lambda G^{\frac{1}{2}}Q(\lambda,\eta)G^{\frac{1}{2}}
$$
where 
$$
Q(\lambda,\eta)=-\eta I+\lambda G^{-\frac{1}{2}}MG^{-\frac{1}{2}}-\lambda^{-1}G^{-\frac{1}{2}}AG^{-\frac{1}{2}}.
$$

For $\lambda\not=0$ we note that  if $ker Q(\lambda,\eta)\not=\{0\}$ then its dimension is the geometric multiplicity of the eigenvalue $\lambda$ of
 the pencil $Q(\cdot,\eta)$ as well as the geometric multiplicity of the eigenvalue $\eta$ of the pencil $Q(\lambda,\cdot)$. The algebraic multiplicities can
 be different, but for $\lambda\in \R$, we have a standard problem for a self-adjoint operator  with the spectral parameter $\eta$, and hence  all
 eigenvalues of $Q(\lambda,\cdot)$ for real $\lambda$ are real and semisimple.     Thus, we have 

\vspace*{3mm} {\bf Lemma 2.11} (cf. \cite{MP}, Lemma 1.2.10) {\it Assume that  $G>>0$, let
 $\lambda_0\in \R\backslash \{0\}$ and let $\eta_0$ be an eigenvalue of the pencil $Q(\lambda_0,\cdot)$ with (geometric) multiplicity 1.  Then  there are $\epsilon>0$ and $l$ real analytic functions

\begin{equation}
\label{2.17}
\eta_s(\lambda)=\eta_0+\mathop{\sum}\limits_{k=p_s}^{\infty}b_{s,k}(\lambda-
\lambda_0)^k, \ \  s=1,2,...,l, \ \ |\lambda-\lambda_0|<\epsilon, 
\end{equation} 
where $p_s\in\N$,
$b_{sp_s}\in\R\backslash\{0\}$, $b_{sk}\in\R$ for $k>p_s$, such that $(\eta_s(\lambda))_{s=1}^l$ represents the eigenvalues near $\eta_0$ of
 the pencil $Q(\lambda,\cdot)$, counted with multiplicity, for each $\lambda\in\C$ with $|\lambda-\lambda_0|<\epsilon$.
}

\proof For real $\lambda$, the eigenvalues of the self-adjoint operator function $Q(\lambda,\cdot)$ are real. Hence the lemma immediately follows from Theorem 2.8 and Remark 2.9.
 \eproof

\vspace*{3mm} {\bf Lemma 2.12}  (\cite{MP}, Lemma 1.4.1) {\it Let the conditions of Lemma 2. 11
be satisfied. Let $\lambda_0>0$ and
$\eta_0>0$, then in some neighborhood of ($\lambda_0, \eta_0$),
i.e. in $\{(\lambda,\eta):|\lambda-\lambda_0|<\epsilon,
|\eta-\eta_0|<\delta,\epsilon>0,\delta>0\}$ all the eigenvalues
are given by the following formula }
\begin{equation}
\label{2.18}
\lambda_j(\eta)=\lambda_0+\mathop{\sum}\limits_{k=1}^{\infty}\beta_k
((\eta-\eta_0)^{\frac{1}{r}}_j)^k, \ \ (j=1,2,...,r),
\end{equation}
{\it where $\beta_1\not=0$ is  real if $r$ is odd and real or pure imaginary if $r$ is even,
$(\eta-\eta_0)^{\frac{1}{r}}_j$, ($j=1,2,...,r$) means the
complete set of branches of the root}.

\proof We obtain this result immediately after inverting
(\ref{2.17}). \eproof


\vspace*{3mm} {\bf Theorem 2.13} {\it Let, in addition to Condition
I, $M>>0$ and $G>>0$  then for negative eigenvalues
$\lambda_{-j}$  of $L$ 
there exists a subsequence of  positive eigenvalue (denote them $\lambda_{j}$ (
$\lambda_{j+1}\geq\lambda_j$ ))
such that}
\begin{equation}
\label{*****}
 \lambda_j+\lambda_{-j}\geq 0.
\end{equation}
 \proof 
We index the eigenvalues in such a way that if $\lambda_{-j}(0)<0$ then $\lambda_{j}\left(0\right)=-\lambda_{-j}(0)$ and
 $\lambda^2_{j}(0)\leq \lambda^2_{j+1}(0)$.

 The eigenvalues of $L(\cdot,\eta)$ are piecewise
analytic functions of $\eta$. The eigenvalues  may loose analyticity only when
they collide.  This follows from the results above. 

The
eigenvalues located on $(-\infty, 0)$   are analytic functions of
$\eta>0$ because there are no nonreal eigenvalues in the open left half-plane (see Lemma 2.4, Statement 2 ) and, consequently, for all $\eta>0$
 and $\lambda\in (-\infty, 0)$ the collision happens according to  (\ref{2.18}) with $r=1$ and $\beta_1\in \R\backslash \{0\}$.

Differentiating $L(\lambda_j(\eta),\eta)y_j(\eta)=0$ with respect to $\eta$ and taking the inner product with $y_j(\eta)$  leads to
$$
\lambda_j^{\prime}(\eta)((2\lambda_j(\eta)M-G)y_j(\eta),y_j(\eta))-\lambda_j(\eta)(Gy_j(\eta),y_j(\eta))=0
$$
and therefore
\begin{equation}
\label{2.20}
\lambda_{j}^{\prime}(\eta)=\frac{\lambda_{j}(\eta)\left(Gy_{j}(\eta),
y_{j}(\eta)\right)}{2\lambda_{j}\left(\eta\right)
(My_{j}(\eta),y_j(\eta))-\eta\left(
Gy_{j}\left(\eta\right),y_{j}\left(\eta\right)\right)}.
\end{equation}
Here $(My_j(\eta),y_j(\eta))$ and $(My_j(\eta),y_j(\eta))$ are positive and depend continuously on $\eta$. 
It follows from (\ref{2.20}) that if $\lambda_j(\eta)<0$ then  $\lambda_j^{\prime}(\eta)>0$. These negative eigenvalues do not cross the 
origin due to Lemma 2.7.

The denominator of (\ref{2.20}) is  positive for sufficiently small $\eta>0$ and $\lambda_j(\eta)>0$ while that the numerator is nonnegative.
 We therefore have shown that $\lambda_j(\eta)$ and $\lambda_{-j}(\eta)$ are nondecreasing, which gives  
$$
 \lambda_j(\eta)+\lambda_{-j}(\eta)\geq 0
$$
for $\eta\geq 0$ small enough.

While $\eta>0$ increases, $\lambda_{j}^\prime\left(\eta\right)$
can change its sign only when the denominator in the right-hand
side of (\ref{2.20}) vanishes, i.e. when eigenvalues collide. If
such a collision takes place on the interval $\left(0
,\infty\right)$, then the eigenvalues involved behave according
to formula (\ref{2.18}). Such a coalescence on the half-axis $(0
,\infty)$ is of one of the following three kinds.

 The first one
is the case of $r$ odd in (\ref{2.18}) and $\beta_1>0$  real. In this case we identify the
eigenvalue moving to the right  along the real axis after the
collision with the one which moved to the right along the real
axis before the collision.

 By a collision of the second kind
we mean one which has $r$ even and $\beta_1\in\R\backslash\{0\}$  in (\ref{2.18}). After such a collision two
new real eigenvalues appear which are moving in 
opposite directions along the real axis, and such a
collision cannot violate Theorem 2.13.

 The third kind of
collision has even $r$ and pure imaginary $\beta_1\not=0$. Let
$\lambda_{j}(\eta)$ take part in such a collision at
$\eta=\eta_0\in(0,1]$. Then a collision of the second kind
indeed occurred at some $\eta\in\left(0,\eta_0,\right)$ in some
point $\lambda_{\times}\in\left(\lambda_{j}(\eta_0),\infty\right)$ on
the real  axis. In this case the eigenvalue that has arisen
after this collision and is moving to the right is identified as
$\lambda_{j}(\eta)$  \eproof

{\bf Remark 2.14} The eigenvalues for which the denominator \\  $2\lambda_{j}\left(\eta\right)
(My_{j}\left(\eta\right),y_j(\eta))-\eta\left(
Gy_{j}\left(\eta\right),y_{j}\left(\eta\right)\right)$ is positive (negative) are called eigenvalues of type I (type II) in \cite{KO}. 
 However,  we use terms  'type I' and  'type II'  in Section 3 differently.   

{\bf Lemma 2.15} {\it Let $M\geq m I$, $m>0$, then 
for all $\eta\in [0,1]$ the
nonreal  eigenvalues (if any) lie in a bounded  domain 
$\{\lambda: 0\leq{\rm Re}\leq \frac {\eta}{2m}||G^{\frac{1}{2}}||^2\}$, 
$|{\rm Im}\lambda|\leq m^{-\frac{1}{2}}\beta^{\frac{1}{2}}$. }

\proof
 Let $\lambda\notin\R$ be an
eigenvalue of $L(\cdot,\eta)$ and $y$ be the corresponding
eigenvector. Then
$$
\lambda^2(My,y)-\eta\lambda(Gy,y)-(Ay,y)=\lambda^2(My,y)-\eta\lambda||G^{\frac{1}{2}}y||^2-(Ay,y)=0
$$
and, since $\lambda$ is not real,
$$
{\rm Re}\lambda=\frac{1}{2}\eta||G^{\frac{1}{2}}y||^2(My,y)^{-1}, \ \ {\rm Im}\lambda=\frac{1}{2}(My,y)^{-1}\sqrt{|\eta^2||G^{\frac{1}{2}}y||^4+4(My,y)(Ay.y)|}
$$
what implies the statement of the lemma.
\eproof

In what follows writing {\it number of eigenvalues in a domain} we mean the number with account of their algebraic multiplicities.

{\bf Theorem 2.16} {\it Let the conditions of Theorem 2.13 be satisfied.   If the negative eigenvalues $\lambda_{-j}$ are arranged in pairs
 with positive eigenvalues $\lambda_j$ such that (\ref{*****}) is true then the number of positive 
eigenvalues not included in the pairs is
$$
2\kappa_A- \kappa_c,
$$
where $\kappa_A$ is the number of 
the negative eigenvalues of the
 operator $A$ and $\kappa_c$ is the number nonreal eigenvalues.}  

\proof
According to Theorem 2.8 the total multiplicity is preserved locally. Due to Lemma 2.15 the nonreal  eigenvalues do not disappear at infinity 
while  $\eta$  grows from 0 to 1. For $\eta=0$ we have $2\kappa_A-\kappa_c=0$ and there are no positive eigenvalue not included in a pair with 
a negative eigenvalue.

Thus the number
 of the eigenvalues in the open upper half-plane may reduce only at collisions on the real axis. 
Let $s$ eigenvalues reach a point on the real axis at $\eta=\eta_0$ moving from the open upper half-plane. Due to the symmetry of
 the spectrum with respect to the real axis the same number of eigenvalues reaches the same point moving from the open lower half-plane.   
This results in disappearing of $s$ eigenvalues from each of  open upper and lower half-planes and appearing $2s$ real eigenvalues. \eproof 

\section{Pencils with rank one  linear part}
\setcounter{equation}{0}

Let us consider  the quadratic operator pencil $L(\lambda,\eta)$
with operators $M$, $G$, $A$ acting in the Hilbert space $H\oplus
\C$ and satisfying Condition I. Moreover, in this section  let the following
condition be valid:

{\bf Condition II:}
$$
G=b\left(\matrix{0 & 0 \cr 0 & 1 }\right), \ \ \ b>0.
$$

\vspace*{3mm} {\bf Lemma 3.1} {\it 1.  Let $\lambda>0$ and $- \lambda$   be  eigenvalues of the operator pencil
$L(\cdot,\eta_0)$ for some $\eta_0\in(0,1]$. Then $\lambda$
and $-
\lambda$ are eigenvalues of $L(\cdot,\eta)$ for all
$\eta\in[0,1]$.

2. Let $\lambda=i\tau$ with $\tau\in \R\backslash\{0\}$ be an eigenvalue of the
operator pencil $L(\cdot,\eta_0)$, where $\eta_0\in(0,1]$. Then
$i\tau$ and $-i\tau$ are eigenvalues of $L(\cdot,\eta)$ for
all $\eta\in[0,1]$. }

\proof  1. Let $y_1=\left(\matrix{y_{11}\cr y_{12}}\right)$ be an eigenvector of $L(\cdot,\eta_0)$ corresponding
 to the eigenvalue $\lambda$ and let $y_2=\left(\matrix{y_{21}\cr y_{22}}\right)$ be an eigenvector of $L(\cdot,\eta_0)$
corresponding to $-\lambda$. Then 
\begin{equation}
\label{*}
(\lambda^2M-\lambda\eta_0G-A)y_1=0,
\end{equation}
\begin{equation}
\label{**}
(\lambda^2M+\lambda\eta_0G-A)y_2=0,
\end{equation}
and consequently
$$
\lambda^2(y_2,My_1)-\lambda\eta_0(y_2, Gy_1)-(y_2,Ay_1)=0,
$$
$$
\lambda^2(My_2,y_1)+\lambda\eta_0(Gy_2,y_1)-(Ay_2,y_1)=0.
$$
Taking into account that $M$, $G$ and $A$ are selfadjoint, the difference of the above equations gives
$$
0=(Gy_2,y_1)=by_{22}\overline{y}_{12}.
$$ 
Then one of the factors  must be zero, say $y_{12}=0$, which gives $Gy_1=0$. Hence (\ref{*}) and (\ref{**}) lead to $L(\pm\lambda, \eta)y_1=0$, which completes the proof of part 1. 

2. Due to the symmetry of the spectrum, see Lemma  2.3  part 2, it follows that if $i\tau$ ($\tau\in\R\backslash \{0\}$) is an eigenvalue
 of $L(\cdot,\eta_0)$, then also $-i\tau$ is an eigenvalue $L(\cdot,\eta_0)$. Let $y$ be an eigenvector of $L(\cdot,\eta_0)$ corresponding to the eigenvalue $i\tau$. Then
$$
-\tau^2(My,y)-i\tau\eta_0(Gy,y)-(Ay,y)=0.
$$
Since $M$, $G$ and $A$ are selfadjoint, all three  inner products are real. Therefore $\lambda\not=0$ and $\eta_0>0$ give $(Gy,y)=0$ and thus $Gy=0$ because  $G\geq 0$. 
It follows $L(\pm\lambda,\eta)y=L(\lambda,\eta_0)y=0$ for all $\eta\in [0,1]$.
\eproof 


\vspace*{3mm}

{\bf Definition 3.2} (see Definition 1.5.2 in \cite{MP}) {\it Let  $\eta_0\in (0,1]$ and
 $m_I(\lambda):=\mathop{min}\limits_{\eta\in (0,1]}m(\lambda,\eta)$, where $m(\lambda,\eta)$ denotes the multiplicity of an eigenvalue $\lambda$ of the pencil $L(\cdot,\eta)$. \\
 1.
An eigenvalue $\lambda$ of $L(
\cdot,\eta_0)$  is said to be an eigenvalue of type I if $\lambda$ is an eigenvalue of the pencil $L(
\cdot,\eta)$ for each $\eta\in (0,1]$. \\
2. For $\lambda\in\C$ let $m_0(\lambda)=dim (ker L(\lambda)\cap ker G)$. If $m_0(\lambda)>0$, then each nonzero vector in $ker L(\lambda)\cap ker G$ is called an eigenvector of type  I for $L$ at $\lambda$. \\
3.  An  eigenvalue $\lambda$ of the pencil $L(\cdot,\eta_0)$
is said to be an eigenvalue of type II $m(\lambda,\eta_0)\not=m_I(\lambda)$. }

\vspace*{3mm}

{\bf Remark 3.3} 1.  An eigenvalue can be both of type I and type II. If this is the case for some $\eta$, then we say that $\lambda$
 is an eigenvalue of the pencil $L(\cdot, \eta)$ of type I multiplicity $m_I(\lambda)$  and of type II multiplicity $m(\lambda,\eta)-m_I(\lambda)$.\\
2. If $0$ is an eigenvalue of the pencil $L$, then it follows from Lemma  2.7 that $0$ is an eigenvalue of $L(\cdot,\eta)$ for all $\eta\in(0,1]$,
 and, if $dim ker A=n$, the algebraic multiplicity $m(0,\eta)$ is $2n$ if  $Gy=0$ for any $y\in ker A$ and $2n-1$ if  there exists $0\not=y\in ker A$ such that $Gy\not=0$.\\
3. If $ker M\cap ker A=\{0\}$, then the pencil $L(\cdot,0)$ satisfies Condition I. Since eigenvalues $\lambda$ of type I are eigenvalues 
of the pencil $L(\cdot,\eta)$ for all $\eta\in(0,1]$, it follows from (\ref{2.18}) that $m_I(\lambda)$ branches of the eigenvalue
 $\lambda$ are constant near $\eta=0$, so that $m_I(\lambda)\leq m(\lambda,0)$, whereas the remaining $m(\lambda,0)-m_I(\lambda)$ branches are not constant.

{\bf Lemma 3.4} 
{\it Assume that $ker M\cap ker A=\{0\}$. Then the eigenvalues of type I of $L(\cdot,\eta)$, which are independent of $\eta\in (0,1]$, are located on imaginary and real axes  symmetrically with respect to $0$. If additionally 
$M+G\geq\epsilon I$ ($\epsilon>0$), at most finitely many of the eigenvalues of type I are on the imaginary axis.}

The proofs of this and the next lemmas are the same as the proofs of Lemmas 1.5.4 and 1.5.5  in \cite{MP}, respectively,  with the only change $K$ for $G$.

\vspace*{3mm} {\bf Lemma 3.5} {\it 1.  For all $\lambda, \eta\in\C$, $ker L(\lambda,\eta)\cap ker G=ker L(\lambda)\cap ker G$, that is
$ker L(\lambda,\eta)\cap ker G$ is independent of $\eta$. In particular, $m_0(\lambda)\leq m_I(\lambda)$.\\
2. Let $\lambda\not=0$, $\eta\in (0,1]$ and assume that $ker L(\lambda)\cap ker G\not=\{0\}$. Then no eigenvector $y_0\in ker L(\lambda\cap ker G=ker L(\lambda)\cap ker G $ of $L(\cdot,\eta)$ at $\lambda$ has an associated vector.\\
3. If $ker M\cap ker A=\{0\}$, then $m_I(\lambda)=m_0(\lambda)$ for all $\lambda\in\C\backslash \{0\}$.}

{\bf Theorem 3.6} {\it Assume that  $ker M\cap ker A=\{0\}$.\\
1. $\lambda\not=0$ is an 
eigenvalue of type I of the operator pencil $L$ if and only if $\lambda$ is an eigenvalue of the pencil $L(\cdot,0)$  having an eigenvector of the form $(y_0,0)^T$, and $m_I(\lambda)$ is the dimension of the space of eigenvectors of this form. \\
2. If $\lambda\not=0$ is an eigenvalue of type I of the pencil $L$, but not an eigenvalue of  type II, then $\lambda$ is semisimple.\\
3.  If $\lambda\not=0$ is an eigenvalue  of the pencil $L$ of type II, then $ker L(\lambda)$ has a basis consisting of $m_I(\lambda)$ eigenvectors of type I
 and one eigenvector $y_0$ with $Gy_0\not=0$ with maximal chain length $m(\lambda)-m_I(\lambda)$, that is, there is a chain $(y_j)_{j=0}^{m(\lambda)-m_I(\lambda)-1}$ 
of the eigenvector $y_0$ and, if $m(\lambda)-m_I(\lambda)>1$, associated vectors of $L$ at $\lambda$.\\  
4. If  $\lambda\not=0$ is an eigenvalue  of type II of the pencil $L$, then $-\lambda$ is not an eigenvalue of type II of the pencil $L$.}

The proof of Theorem 3.6  is the same as the proof of Theorem 1.5.6 in  \cite{MP} with the change of $K$ for $G$.

Let us describe location of the eigenvalues $\{\lambda_j^{II}(\eta)\}$ of type II.
Let us enumerate them as follows:
$$
(\lambda^{II}_{1}(0))^2<(\lambda^{II}_2(0))^2<...<(\lambda^{II}_{\kappa_{II}}(0))^2<0\leq (\lambda^{II}_{\kappa_{II}+1}(0))^2<(\lambda^{II}_{\kappa_{II}+2}(0))^2<... 
$$
$$
  \lambda^{II}_{-j}(0)=-\lambda^{II}_{j}(0).
$$
Here $2\kappa_{II}$ is the number of pure imaginary eigenvalues of $L(\cdot, 0)$ of type II.

{\bf Proposition 3.7} {\it The negative eigenvalues of type II preserve their order:}
$$
...<\lambda^{II}_{-j-1}(\eta)<\lambda^{II}_{-j}(\eta)<...<\lambda^{II}_{-\kappa_{II}-1}(\eta)\leq 0.
$$ 
\proof We know from  Lemma 2.5 that all the negative eigenvalues are semisimple. From Statement 3 of Theorem 3.6 we conclude that the eigenvalues of type II are 
geometrically simple (however, an eigenvalue of type II can for some value of $\eta$ coincide with an eigenvalue of type I).
Since $\lambda_j^{II}(\eta)$ are continuous functions of $\eta$ we arrive at the statement of the proposition. 
\eproof

{\bf Theorem 3.8} {\it Assume that $ker M\cap ker A=\{0\}$ and that $M+G>>0$. Then the eigenvalues of type II of the operator pencil  $L$ possess the following properties.\\
1. $L(\cdot, \eta)$ has nonreal eigenvalues of type II  located symmetrically with respect to the real axis. Denote their number by $2\kappa_{II}(\eta)\geq 0$. \\
2. For all $\eta\in (0,1]$ these nonreal eigenvalues of type II  lie in the open right half-plane.  \\ 
3.  $|\lambda_{-j}^{II}(\eta)|\notin\{\lambda^{II}_j(\eta)\}$ for all $j\geq \kappa_{II}(0)+1$ if $\lambda_{ \kappa_{II}(0)+1}>0$ and for  all $j\geq \kappa_{II}(0)+2$ if $\lambda_{ \kappa_{II}(0)+1}(0)=0$. \\
4.  The number $n_j(\eta)$ of eigenvalues of type II in each interval $(|\lambda_{-j}^{II}(\eta)|, |\lambda_{-j-1}^{II}(\eta)|)$ ($j\geq \kappa_{II}(0)+1$ if $\lambda^{II}_{\kappa_{II}(0)+1}(0)>0$ and  $j\geq \kappa_{II}(0)+2$ if $\lambda^{II}_{\kappa_{II}(0)+1}(0)=0$ ) is odd. \\
5. If  $\lambda^{II}_{\kappa_{II}+1}(0)>0$       the number $n_0(\eta)$ of eigenvalues of type II in the interval $(0, |\lambda^{II}_{-\kappa_{II}-1}(\eta)|)$ is even.\\
6.  Denote by 
$$
\tilde{\kappa}_{II}(\eta)=\frac{1}{2}\left\{\begin{array}{c}n_0(\eta)-1+\mathop{\sum}\limits_{j=1}^{\infty}(n_j(\eta)-1) \ \ {\rm if} \ \ \lambda^{II}_{-\kappa_{II}}(0)=0,  \\ n_0(\eta)+\mathop{\sum}\limits_{j=1}^{\infty}(n_j(\eta)-1), \ \ {\rm if} \ \ \lambda^{II}_{-\kappa_{II}}(0)\not=0 \end{array}\right. .
$$
Then $\tilde{\kappa}_{II}(\eta)+\kappa_{II}(\eta)+\kappa_I=\kappa_A$
 where $\kappa_A$ is the number of  negative eigenvalues of the operator pencil $\lambda M-A$ and $2\kappa_I$ is 
the number (independent of $\eta$) of pure imaginary eigenvalues of $L(\cdot, \eta)$ of type I.}

 \proof
Statement 1 follows from Lemma 2.3. 

By Lemma 2.15 we know that the nonreal  eigenvalues lie in the closed right half-plane. But if such an eigenvalue is pure imaginary, then it is of type I. This means 
that nonreal eigenvalues of type II lie in the open right half-plane. Statement 2 is proved.

If $|\lambda_{-j}^{II}(\eta)|\in\{\lambda^{II}_j(\eta)\}$ for some    $j\geq \kappa_{II}(0)+1$ if $\lambda_{ \kappa_{II}(0)+1}>0$ or some $j\geq \kappa_{II}(0)+2$ if $\lambda_{ \kappa_{II}(0)+1}=0$ then by Lemma 3.1 $\lambda^{II}_j(\eta)$ is independent of $\eta$ and therefore is of type II. This contradiction proves Statement 3.

Since the right-hand side of (\ref{2.20}) is nonnegative for small $\eta>0$ and $\lambda_j(\eta)\in\R$ the real eigenvalues for such $\eta$ move to the right along the real axis. Therefore, the number of  eigenvalues of type II  in each interval $(|\lambda_{-j}(\eta)|, |\lambda_{-j-1}(\eta)|)$ ($j\geq \kappa_{II}(0)+1$ if $\lambda^{II}_{\kappa_{II}(0)+1}>0$ and  $j\geq \kappa_{II}(0)+2$ if $\lambda^{II}_{\kappa_{II}(0)+1}=0$ )  is 1 for $\eta>0$ small enough. This  number can increase only when a collision of the second kind of eigenvalues of type II  happens on the real axis.  Such a collision increases the number of the 
eigenvalues in such an interval by 2. Then this number  can increase by 2 as the result of a collision of the second kind and decrease by 2 as the result of a collisions of the third type. Thus $n_j(\eta)$ is odd for all $\eta>0$ what proves Statement 4. 

Let   $\lambda^{II}_{\kappa_{II}+1}(0)>0$. Then for $\eta>0$ small enough there are no eigenvalues of type II on the interval $(0, |\lambda_{-\kappa_{II}(0)-1}|)$. This number can increase by 2 as a result of collision of the second kind and decrease by 2 as a result of collision of the third kind. Statement 5 follows.

It is clear that $\tilde{\kappa}_{II}(0)=0$ and $\kappa_{II}(0)+\kappa_{I}=\kappa_A$. The number $2\kappa_{II}(\eta)$ of nonreal eigenvalues of type II increases (decreases) by 2 at collisoons of the second (third) kind. But simultaneously $\tilde{\kappa}_{II}$ decreases (increases) by 2.  Thus, the sum $\kappa_{II}(\eta)+\tilde{\kappa}_{II}(\eta)$ remains unchanged at collisions of any of the three kinds. This proves Statemens 6.
 \eproof

\section{Applications} \setcounter{equation}{0}

1. We consider a Sturm-Liouville problem  with the gyroscopic condition at the right end
\begin{equation}
\label{4.1}
 -y^{\prime\prime}+q(x)y=\lambda^2y,
 \end{equation}
 \begin{equation}
 \label{4.2}
y(0)=0,
\end{equation}
\begin{equation}
\label{4.3}
 y^{\prime}(a)+\lambda \alpha y(a)=0.
\end{equation}
 Here $\lambda$ is the spectral parameter, the parameter $\alpha>0$ and the potential $q$
is real-valued and belongs to $L_2(0,a)$. 


 Let us introduce the operators $A$, $G$ and $M$ acting in the
Hilbert space $H=L_2(0,a)\cup\C$ according to the formulae:
$$
 A\left(
\begin{array}{c}
v(x) \\
c \\
\end{array}
\right) =\left(
\begin{array}{c}
-v^{\prime\prime}\left(x\right) +q\left( x\right) v\left( x\right) \\
v^{\prime }(a)
\end{array}
\right),
$$
$$
 D\left( A\right) =\left\{
\begin{array}{c}
\left(
\begin{array}{c}
v\left( x\right) \\
c
\end{array}
\right): \ \ v\left( x\right) \in W_{2}^{2}\left( 0,a\right), \ \
\ \ v\left( 0\right) =0, \ \ c=v(a)
\end{array}
\right\},
$$
$$
M=\left(
\begin{array}{cc}
I & 0 \\
0 & 0
\end{array}
\right), \ \ G=\left(
\begin{array}{cc}
0 & 0  \\
0 & \alpha
\end{array}
\right).
$$
The operator  $A$ is selfadjoint and bounded below ($A\geq \beta I$) and
 there exists $-\beta_1<-\beta\leq ||y||^{-2}(Ay,y)$ such that
$(A+\beta_1I)^{-1}$ is a compact operator (see \cite{MP}, Proposition 2.1.1).

We associate   the   quadratic operator pencil 
\begin{equation}
\label{***}
 L\left( \lambda \right) =\lambda ^{2}M-\lambda G-A
\end{equation}
 with problem (\ref{4.1})--(\ref{4.3}). 

We identify the spectrum of problem \rf{4.1} - \rf{4.3} with the
spectrum of the pencil $L\left( \lambda \right)$. It is clear that
$M\geq 0$, $G\geq 0$ and $M+G>>0$.
The spectrum of the pencil consists of normal eigenvalues (see
Section 2 of this paper).

Let us prove that all of these eigenvalues are of type II. Suppose
a pure imaginary $i\tau\not=0$ is an eigenvalue. Then by Lemma  3.1 $-i\tau$ is also an eigenvalue and according
to the proof of Lemma 3.1 $c=v(a)=0$ in (\ref{4.3}). Therefore,
the second component of the equation $L(\lambda_0)Y=0$ gives
$v^{\prime}(a)=0$ what contradicts $v(a)=0$. In the same way, one
can prove that there are no symmetrically located real
eigenvalues and that the possible eigenvalue at the origin is
simple.

 Thus, the conditions of Theorem 3.8 are satisfied and all the eigenvalues are of type
 II and therefore statements 1)--6) of Theorem 3.8 are valid. Thus we obtain

{\bf Theorem 4.1} {\it
1. Problem (\ref{4.1})--(\ref{4.3}) may  have nonreal eigenvalues   located symmetrically with respect to the real axis. Denote their number by $2\kappa\geq 0$. \\
2. All the nonreal eigenvalues  lie in the open right half-plane.  \\ 
3.  If we denote by $\{\lambda_j\}_{-\infty}^{\kappa_0}$ the negative eigenvalues of problem (\ref{4.1})--(\ref{4.3}) then $|\lambda_{k}|\notin\{\lambda_j\}_{-\infty,j\not=0}^{\infty}$ for all $-\infty<k\leq \kappa_0$. \\
4.  The number $n_j$ of eigenvalues  in each interval $(|\lambda_{j}|, |\lambda_{j-1}|)$ ($-\infty< j\leq \kappa_0$) is odd. \\
5. If  $0\notin\{\lambda_{j}\}_{-\infty, j\not=0}^{\infty}$ then       the number $n_0$ of eigenvalues in the interval $(0, |\lambda_{\kappa_0}|$) is even.  If  $0\in\{\lambda_{j}\}_{-\infty, j\not=0}^{\infty}$ then       the number  of eigenvalues in the interval $(0, |\lambda_{\kappa_0}|)$ is odd and $0$ is a simple eigenvalue.\\
6.  Denote by 
$$
\tilde{\kappa}=\frac{1}{2}\left\{\begin{array}{c}n_0+\mathop{\sum}\limits_{j=1}^{\infty}(n_j-1) \ \ {\rm if} \ \ 0\not\in\{\lambda_j\}_{-\infty,j\not=0}^{\infty},  \\ n_0-1+\mathop{\sum}\limits_{j=1}^{\infty}(n_j-1), \ \ {\rm if} \ \ 0\in\{\lambda_j\}_{-\infty, j\not=0}^{\infty} \end{array}\right. .
$$
Then $\tilde{\kappa}+\kappa=\kappa_A$
 where $\kappa_A$ is the number of  negative eigenvalues of the operator pencil $\lambda M-A$ or what is the same of problem}
\begin{equation}
\label{4.11}
 -y^{\prime\prime}+q(x)y=\lambda y,
 \end{equation}
 \begin{equation}
 \label{4.12}
y(0)=y'(a)=0.
\end{equation}

2.  In the second example we have eigenvalues of both types. 
Let $q=const>0$. Consider the following spectral problem
\begin{equation}
\label{4.7}
-y_j^{\prime\prime}+qy_j=\lambda^2y_j, \ \ \  j=1,2,
\end{equation}
\begin{equation}
\label{4.8}
y_1(0)=y_2(0)=0,
\end{equation}
\begin{equation}
\label{4.9}
y_1(a)=y_2(a),
\end{equation}
\begin{equation}
\label{4.10}
y_1^{\prime}(a)=y_2^{\prime}(a)+\lambda\alpha y_2(a).
\end{equation}

 Let us introduce the operators $A$, $G$ and $M$ acting in the
Hilbert space $H=L_2(0,a)\oplus L_2(0,a)\oplus\C$ according to the formulae:
$$
 A\left(
\begin{array}{c}
v_1(x) \\ v_2(x) \\
c 
\end{array}
\right) =\left(
\begin{array}{c}
-v_1^{\prime\prime}\left(x\right) +q v_1\left( x\right) \\ 
-v_2^{\prime\prime}\left(x\right) +q v_2\left( x\right) \\
v_2^{\prime }(a)
\end{array}
\right),
$$
$$
D(A)=
$$
$$
 \left\{
\begin{array}{c}
\left(
\begin{array}{c}
v_1\left( x\right) \\ v_2(x) \\
c
\end{array}
\right): \ \ v_1\left( x\right) \in W_{2}^{2}\left( 0,a\right), \ \ v_2\left( x\right) \in W_{2}^{2}\left( 0,a\right),
\ \ v_1\left( 0\right) = v_2(0)=0, \ \ c=v_2(a)
\end{array}
\right\}, 
$$
$$
M=\left(
\begin{array}{ccc}
I & 0 & 0 \\
0 & I &  0 \\
0 & 0 & 0
\end{array}
\right), \ \ G=\alpha \left(
\begin{array}{ccc}
0 & 0 & 0 \\
0 & 0 & 0 \\
0 & 0 & 1  
\end{array}
\right).
$$

The corresponding operator pencil $L(\lambda)$ given be (\ref{***}) satisfies conditions of Theorem 3.8. Moreover, in case of $q=const$ we can find 
the characteristic function of this problem, i.e. the function the set of zeros of which coincides with the spectrum of  problem (\ref{4.7})--(\ref{4.9}):
\begin{equation}
\label{yyy}
\omega(\lambda)=\frac{\sin\sqrt{\lambda^2+q } \ a}{\sqrt{\lambda^2+q}} \  \left(2\cos \sqrt{\lambda^2+q} \  a+\alpha\lambda\frac{\sin\sqrt{\lambda^2+q} \ a}{\sqrt{\lambda^2+p}}\right).
\end{equation}
The spectrum of problem (\ref{4.7}) -- (\ref{4.10}),  i.e. the set of zeros of $\omega$ consists of two subsequences $\{\lambda_j^I\}_{-\infty, j\not=0}^{\infty}$ and $\{\lambda_j^{II}\}_{-\infty, j\not=0}^{\infty}$ where
\begin{equation}
\label{.}
\lambda_j^{I}=\pm\sqrt{\left(\frac{\pi j}{a}\right)^2-q}, \ \ \  \ \  j=\pm 1, \pm 2, ...
\end{equation}
and  $\{\lambda_j^{II}\}_{-\infty, j\not=0}^{\infty}$ is the set of zeros  of the function $\left(2\cos \sqrt{\lambda^2+q} \  a+\alpha\lambda\frac{\sin\sqrt{\lambda^2+q} \ a}{\sqrt{\lambda^2+p}}\right)$.
According to  Definition 3 2 (with $\eta=\alpha$) we obtain

{\bf Theorem 4.2} {\it The zeros $\{\lambda_j^I\}_{-\infty, j\not=0}^{\infty}$ are eigenvalues of the type I of problem (\ref{4.7})--(\ref{4.10}) while the zeros $\{\lambda_j^{II}\}_{-\infty, j\not=0}^{\infty}$
are the eigenvalues of type II.}

Analyzing the behavior of the function $\left(2\cos \sqrt{\lambda^2+q} \  a+\alpha\lambda\frac{\sin\sqrt{\lambda^2+q} \ a}{\sqrt{\lambda^2+p}}\right)$ we conclude
that (for $q=const$)

(i)  $n_j=1$ for all $j$ and $n_0=0$ if $0\in\{\lambda_j\}_{-\infty, j\not=0}^{\infty}$ and $n_0=1$ if $0\notin\{\lambda_j\}_{-\infty, j\not=0}^{\infty}$. 

(ii)  An eigenvalue of type I cannot coincide with an eigenvalue of type II for any value of $\alpha>0$ because if
\begin{equation}
\label{99}
\frac{\sin\sqrt{\lambda^2+q } \ a}{\sqrt{\lambda^2+q}}=0 
\end{equation}
and
$$
  \left(2\cos \sqrt{\lambda^2+q} \  a+\alpha\lambda\frac{\sin\sqrt{\lambda^2+q} \ a}{\sqrt{\lambda^2+p}}\right)=0
$$
then
$$
\cos \sqrt{\lambda^2+q} \  a=0
$$
what contradicts (\ref{99}).

(iii) In the same way we conclude that $\{-\lambda^{II}_j\}_{-\infty, j\not=0}^{\infty}\cap \{\lambda_j^{I}\}_{-\infty, j\not=0}^{\infty}=\emptyset$.

To avoid considering many cases we give explicit result for one particular case. Here we describe  the eigenvalues $\{\lambda_j\}_{-\infty, j\not=0}^{\infty}=\{\lambda_j^{I}\}_{-\infty, j\not=0}^{\infty}\cup\{\lambda_j^{II}\}_{-\infty, j\not=0}^{\infty}$ identifying 
$\lambda_{2j-1}=\lambda_J^{II}$ and $\lambda_{2j}=\lambda_j^{I}$ ($j\in\N$). 

Using (i)--(iii) and Theorem 3.8 we obtain

{\bf Corollary 4.3} {\it
 Let $q>0$ and $q^{1\over 2}a\pi^{-1}\in\N$. Then

a) $\lambda=0$ is an eigenvalue of geometric multiplicity 1 and algebraic multiplicity 2;

b) there are $2(q^{1\over 2}a\pi^{-1}-1)$ simple pure imaginary (nonzero) eigenvalues located symmetrically with respect to the origin which can be found by (\ref{.}) with $j=\pm 1, \pm 2, ..., \pm (q^{1\over 2}a \pi^{-1}-1)$;

c)  there are $2q^{1\over 2}a\pi^{-1}$ complex (neither real nor pure imaginary) eigenvalues located symmetrically with respect to the real axis in the open right half plane;

d) real eigenvalues can be indexed such that $\lambda_j<0$ for $-\infty<j\leq -2q^{1\over 2}a\pi^{-1}-1:=\kappa_0$ and the interval $(0,|\lambda_{\kappa_0}|)$ is free of eigenvalues

e)  If we denote by $\{\lambda_j\}_{-\infty}^{\kappa_0}$ the negative eigenvalues of problem (\ref{4.7})--(\ref{4.10}) then $|\lambda_{2k-1}|\notin\{\lambda_j\}_{-\infty,j\not=0}^{\infty}$ for all $-\infty<2k-1\leq \kappa_0$;

f)  The number of eigenvalues  in each interval $(|\lambda_{2j-1}|, |\lambda_{2j-3}|)$ ($-\infty<2j -1\leq \kappa_0$ is 2 (one of type I and one of type II);

g) The number of eigenvalues in each the intervals $(-((\pi(j+1)a^{-1})^2-q)^{1\over 2}, -((\pi j a^{-1})^2-q)^{1\over 2})$ and
 in each the intervals $(((\pi ja^{-1})^2-q)^{1\over 2}, ((\pi (j+1) a^{-1})^2-q)^{1\over 2})$ 
for $j=q^{1\over 2}a\pi^{-1}, q^{1\over 2}a\pi^{-1}+1,...$ is 1.}










South-Ukrainian National Pedagogical University, \\
Staroportofrankovskaya str. 26, Odesa, Ukraine, 65020 \\
e-mail address: vpivovarchik@gmail com  \\
boykohelga@gmail.com


\begin{thebibliography}{99}
\itemsep=0pt
%


\bibitem{DD} D.B. DeBra, R. H. Delp, Rigid body attitude stability and natural frequences in a circular orbit. J. Astronaut. Sci. {\bf 8}  (1961), 14--17. 
%
\bibitem{Fe} V.I. Feodosiev. Vibrations and stability of a pipe conveying a flowing liquid. Inzh. sbornik {\bf 10} (1951), 169--170 [in Russian]
%
\bibitem{GK1}
I.C. Gohberg, M.G. Krein. Introduction to the Theory of Linear
Nonselfadjoint Operators, Transl.  Math.
Monographs, {\bf 18} AMS, New York, 1969.
%
\bibitem {GGK}
I.C. Gohberg, S. Goldberg, M.A. Kaashoek. Classes of Linear
Operators, I. Operator Theory: Adv. Appl.  {\bf 49}
Birkh\"auser, Basel, 1990.
%
\bibitem{GS}
I.C. Gohberg, E.I. Sigal. An Operator generalization of the
logarithmic residue theorem and Rouch\'e's theorem.  Math. USSR
Sb. {\bf 13:1}  (1971),  603-625 [in Russian].
%
\bibitem{IN} Ch. G. Ibadzade, I.M. Nabiev. Recovering of the Sturm-Liouville operator with non-separated  boundary condition dependent of the spectral parameter. to be published in Ukrainian Math. J. [in Russian]
\bibitem{KO}
%
A.G. Kostyuchenko, M.B. Orazov. The problem of oscillations of an
elastic half- cylinder and related self-adjoint quadratic pencils.
 Trudy Sem. Petrovskogo, {\bf 6} (1981), 97-147 [in Russian].
%
\bibitem{M}
A.S. Markus. Introduction to the Theory of Polynomial Operator
pencils, Transl. Math. Monographs {\bf 71} AMS, New York,
1988.
%
\bibitem{Mh}
A.I. Markushevich. Theory of  Functions of a Complex Variable. III, revised English edition, Prentice-Hall Inc.,  Englewood  Cliffs, New York, 1967.
%
\bibitem{Mi1} A.I. Miloslavskii. On stability of linear pipes, Dinamika Sistem, Nesuschih podvizhnuyu raspredelennuyu nagruzku, Collected papers, {\bf 192} Kharkov Aviation Institute, Kharkov, (1980) 34--47 [in Russian]. 
%
\bibitem{Mi2} A.I. Miloslavskii. On the instability spectrum of an operator pencil, Matem. Zametki  {\bf 49} (1991), no. 4, 88--94 [in Russian]; English transl. in: Mathem. Notes,  {\bf 49} (1991), no. 3--4, 391--395.  
%
\bibitem{Mo} A.A. Movchan .On a problem of stability of a pipe with a fluid flowing through it, Prikl. Mat. Mekh. {\bf 29} (1965), 760--762 [in Russian]; English transl.: PMM, J. Appl. Math. Mech. {\bf 29} (1965), 902-904.
%
\bibitem{MP} M. M\" oller M, V. Pivovarchik, Spectral Theory of Operator Pencils, Hermite-Biehler Functions, and Their Applications.
Birkh\"auser, Cham, 2015
%
\bibitem{PI} M.P. Paidoussis, N.T. Issid. Dynamic stability of pipes conveying fluid. J. Sound Vibrat.  {\bf 33} (1974), no. 3, 267--294.
%
\bibitem{P3}
V.N. Pivovarchik. On eigenvalues of a quadratic operator pencil, 
Funkts. Anal. Prilozhen. {\bf 25} (1989),
 80-81 [in
Russian].
%
\bibitem{Pumz}  V. Pivovarchik. On spectra of a certain class of quadratic operator pencils with one-dimensional linear part. Ukrainian Math. J. {\bf 59} (2007), No.5, 702-717. 
%
\bibitem{P6}
V.N. Pivovarchik. On spectra of quadratic operator pencils in the
right half-plane.  Matem. Zametki {\bf 45} (1989), No.6, 101-103 [in Russian].
%
\bibitem{P7}
V.N. Pivovarchik. On the total algebraic multiplicity of the
spectrum in the right half-plane for a class of quadratic operator
pencils, Algebra i Analys {\bf 3} (1991),  No. 2, 223 - 230 [in Russian]; English transl.:  St. Peterburg Math. J.  {\bf 3} (1992),  No. 2.,
447-454.
%
\bibitem{PW}  V. Pivovarchik, H. Woracek.  Shifted Hermite-Biehler functions and their applications, Integral Equations and Operator Theory {\bf 57} (2007) 101-126.
%
\bibitem{Sh} A.A. Shkalikov. Elliptic equations in Hilbert space and associated spectral problems.
 Trudy Sem. Petrovskogo  {\bf 14} (1989) 140-224  [in Russian].
%

\bibitem{Th} W. T. Thompson. Spin stabilization of attitude against gravity torque, J. Austronaut. Sci.  {\bf  9}  (1962), no. 1, 31 - 33.
%

\bibitem{Tya} M. Tyaglov. Self-interlacing polynomials II: matrices with self-interlacing spectrum. Electron. J. Linear Algebra {\bf 32} (2017), 51--57.
%
%
\bibitem{Za} E.E. Zajac. The Kelvin-Tait-Chetaev theorem and extensions. J. Austronaut. Sci.   {\bf 11}  (1964), 46 - 49.


%
\end{thebibliography}
\end{document}